\documentclass[12pt,titlepage]{article}
\usepackage{amsmath}
\usepackage{amssymb}
\usepackage{graphicx}
\usepackage{caption2}
\usepackage{amsfonts}
\usepackage{cite}

\newcommand{\be}{\begin{equation}}
\newcommand{\ee}{\end{equation}}
\newcommand{\bea}{\begin{eqnarray}}
\newcommand{\eea}{\end{eqnarray}}
\newcommand{\bef}{\begin{figure}}
\newcommand{\ef}{\end{figure}}
\newcommand{\bt}{\begin{tabular}}
\newcommand{\et}{\end{tabular}}
\newcommand{\bno}{\begin{enumerate}}
\newcommand{\eno}{\end{enumerate}}

\def\3{\ss}

\catcode`\"=\active
\def"{\accent'177}
\begin{document}

\begin{center}
{\bf\large  Two kinds of  peaked solitary waves of the KdV, BBM \\ and Boussinesq equations}

\vspace{0.5cm} 

Shijun Liao\\
\vspace{0.5cm} 
Dept. of Mathematics \&   State Key Laboratory of Ocean Engineering \\
School of Naval Architecture, Ocean and Civil Engineering\\
Shanghai Jiao Tong University, Shanghai 200240, China\\
\vspace{0.35cm}
(email: sjliao@sjtu.edu.cn)

\end{center}

\hspace{-0.75cm}{\bf Abstract} {\em It is well-known that the celebrated Camassa-Holm equation has the peaked solitary waves,  which have been not reported  for other mainstream models of shallow water waves.    In this letter, the closed-form solutions of peaked solitary waves of the KdV equation,  the BBM equation  and the Boussinesq  equation  are  given for the first time.   All  of  them have either a peakon or an anti-peakon.     Each of them exactly  satisfies the corresponding  Rankine-Hogoniot  jump  condition and should be  understood as weak solution.    Therefore, the peaked solitary waves  might  be  common for most of shallow water wave models,  no matter whether  or not  they are integrable and/or  admit breaking-wave solutions.        
}

%\hspace{-0.75cm} {\bf PACS Number}: 47.35.Bb  

\hspace{-0.75cm}  {\bf Key Words} Solitary waves, peaked crest,  progressive,  discontinuity, weak solution  

\section{Introduction}

Since the solitary surface wave was discovered by John Scott Russell \cite{Russell1844} in 1834,   many models of solitary waves in shallow water  have been developed, such as the  Boussinesq  equation \cite{Boussinesq1872},  the Korteweg  \&  de Vries (KdV) equation \cite{KdV},  the  Benjamin-Bona-Mahony   (BBM)  equation \cite{Benjamin1972},  the Camassa-Holm (CH) equation \cite{Camassa1993PRL}, and so on.  Unlike other models, the  celebrated  CH equation
\begin{equation} 
 u_t + 2 \kappa  u_x - u_{xxt} + 3 u u_x = 2 u_x u_{xx} + u u_{xxx}    \label{geq:CH}
 \end{equation}
 possesses both of the global solutions in time and the wave-breaking solutions,  where $u$ denotes wave elevation,  $x,t$ are the spatial and temporal variables,    the subscript denotes the differentiation,  and $0\leq \kappa \leq 1/2$ is a constant related to the critical shallow water wave speed, respectively.   Especially, when $\kappa = 0$, the CH equation (\ref{geq:CH}) has 
a peaked solitary wave
$ u(x,t)  = c \; \exp(-|x - c\; t|)$, 
 where $c$ denotes the phase speed.    It should be emphasized that, at the wave crest  $x = c \; t$,  the above closed-form solution has no continuous derivatives respect to $x$, thus it does not satisfy Eq. (\ref{geq:CH}) (when $\kappa = 0$) at the crest  $x = c \; t$.  So,  it is not a {\em strong} solution.     However,   Constantin and Molinet \cite{Constantin2000-B} pointed out that $ u(x,t)  = c \; \exp(-|x - c\; t|)$  could be understood   as a weak solution of the CH equation  (\ref{geq:CH}) when $\kappa=0$. 
 
Note that  such kind of discontinuity (or  singularity) exists  widely in natural phenomena, such as dam breaking \cite{Zoppou2000AMM}  in hydrodynamics, shock waves in aerodynamics,  black holes described by the general relativity, and so on.  In the frame of water waves,  Stokes \cite{Stokes1894} found that the limiting gravity wave has a sharp corner at the crest.  In case of the dam breaking, there exist sharp corners of wave elevation at the beginning $t=0$, and such kind of discontinuity of the derivative of wave elevation does not disappear due to the neglect of the viscosity, as shown by Wu and Cheung \cite{Wu2008IJNMF}.     In fact,  problems related to such kind of discontinuity  belong to the so-called  Riemann problem \cite{Wu2008IJNMF, Bernetti2008JCP,  Rosatti2010JCP},  which  is a classic field of fluid mechanics.    

However,  such kind of peaked solitary waves have {\em never}  been  found for other mainstream models of shallow water waves.    Solitary wave solutions  in shallow water with a fixed speed $c$ and permanent form were found  by  Korteweg  \&  de Vries  \cite{KdV} using the KdV equation
\begin{equation}
\zeta_t + \zeta_{xxx} + 6 \zeta \; \zeta_x = 0, \label{geq:KdV:original}
\end{equation}
subject to the boundary conditions
\begin{equation}
\zeta\to 0, \zeta_x\to 0, \zeta_{xx}\to 0, \mbox{as $|x|\to +\infty$},  \label{bc:infinity:original}
\end{equation}
where $\zeta$ denotes the wave elevation.   The KdV equation (\ref{geq:KdV:original})  admits solitary waves but  breaking ones.   The traditional solitary wave of  KdV equation (\ref{geq:KdV:original}) reads
\begin{equation}
\zeta(x,t) = \frac{c}{2} \;  \mbox{sech}^2 \left[\frac{\sqrt{c}}{2} (x-c \; t - x_0)  \right],\label{zeta:traditional}
\end{equation}
where sech denotes a hyperbolic secant function and $x_0$ may be any a constant.  This solitary wave has a smooth crest with always positive elevation $\zeta(x,t)\geq 0$, and besides its phase speed $c$ is dependent upon the wave height, i.e.  $c = 2\zeta_{max}$, so that higher solitary waves propagate faster.  To the best of  the author's knowledge,  no peaked solitary waves of KdV equation (\ref{geq:KdV:original}) were reported.    

The KdV equation (\ref{geq:KdV:original}) is an approximation of the fully nonlinear wave equations that  admit the limiting gravity wave with a peaked crest,   as  pointed out by Stokes \cite{Stokes1894}.    Besides,  as mentioned before, this kind of discontinuity of wave elevation widely exists in hydrodynamic problems such as dam break \cite{Zoppou2000AMM}.     In addition, current investigations \cite{Constantin2000, Dullin2001}  reveal the close relationships  between the CH equation (\ref{geq:CH}) and the KdV equation (\ref{geq:KdV:original}).     Therefore,   one has many reasons to   assume  that  the  progressive solitary waves  of  the  KdV equation (\ref{geq:KdV:original})  might admit  peaked solitary waves, too.       

\section{Peaked solitary waves of KdV equation}%

Write $\xi = x - c \; t - x_0$ and $\eta(\xi)=\zeta(x,t)$.  The original KdV equation (\ref{geq:KdV:original}) becomes
\begin{equation}
-c \eta' + \eta'''+6 \eta \eta' =0, \label{geq:KdV}
\end{equation}
subject to the boundary condition
\begin{equation}
\eta \to0, \;\; \eta'\to 0, \;\; \eta''\to 0, \mbox{as $|\xi|\to +\infty$}, \label{bc:infinity}
\end{equation}
where the prime denotes the differentiation with respect to $\xi$. Besides, there exists the symmetry condition
\begin{equation}
\eta(-\xi) = \eta(\xi), \hspace{1.0cm} \xi \in(-\infty,+\infty)  \label{bc:symmetry}
\end{equation}
so that we need only consider the solution in the interval $\xi \geq 0$. 
Integrating (\ref{geq:KdV}) with (\ref{bc:infinity})  gives
\begin{equation}
-c \eta + \eta'' + 3 \eta^2 = 0, \hspace{1.0cm} \xi \geq 0 .  \label{geq:KdV:1}
\end{equation}
Multiplying it  by $2 \eta'$ and then integrating it with (\ref{bc:infinity}), we have
\begin{equation}
 \eta'^2 = \eta^2 (c-2\eta),   \hspace{1.0cm} \xi \geq 0,   \label{geq:KdV:2}
 \end{equation}
which has real solutions only when
\begin{equation}
\eta  \leq  \frac{c}{2}.  \label{eta:limit}
\end{equation}
Under the restriction (\ref{eta:limit}), we have from (\ref{geq:KdV:2}) that
\begin{equation}
\eta' = \pm \eta \; \sqrt{c-2\eta},  \label{geq:KdV:3}
\end{equation}
say,
\begin{equation}
\frac{d \eta}{\eta \; \sqrt{c-2\eta}} = \pm d \xi, \;\; \;\;  \xi \geq 0. \label{geq:KdV:4}
\end{equation}

Let us first consider the solitary waves with $\eta\geq 0$.   Integrating (\ref{geq:KdV:4}) in case of $0\leq \eta \leq c/2$,   we have
\begin{equation}
\mbox{tanh}^{-1}\left[ \sqrt{1-\frac{2 \eta}{c}} \right] = \pm \frac{\sqrt{c}}{2} \xi + \alpha,
\end{equation}
where $\alpha$  is a constant.  Since $0 \leq 2\eta/c \leq 1$,   the left-hand side of the above equation is non-negative  so that  it  holds for
all  $\xi\geq 0$ if and only if
\[   \mbox{tanh}^{-1}\left[ \sqrt{1-\frac{2 \eta}{c}} \right] = \frac{\sqrt{c}}{2}  \xi  + \alpha, \;\;\;\xi\geq 0,  \alpha \geq 0,     \]
which gives
\begin{equation}
\eta(\xi)  = \frac{c}{2} \; \mbox{sech}^2\left[ \frac{\sqrt{c}}{2} \xi + \alpha \right], \;\;\; \xi\geq 0, \alpha\geq 0.
\end{equation}
Using the symmetry condition (\ref{bc:symmetry}),  it holds 
\begin{equation}
\eta(\xi)  = \frac{c}{2} \; \mbox{sech}^2\left[ \frac{\sqrt{c}}{2} |\xi| + \alpha \right], \;\;\; -\infty<\xi<+\infty, \alpha\geq 0
\end{equation}
in the {\em whole}  interval.    Thus, we have the solitary waves of the first kind
\begin{equation}
\zeta(x,t) = \frac{c}{2}\mbox{sech}^2\left[ \frac{\sqrt{c}}{2} |x-c \;t -x_0|  +  \alpha  \right],  \label{zeta:new:positive}
\end{equation}
where $\alpha \geq 0$,  with the wave height 
\begin{equation}
\zeta_{max} = \left( \frac{c}{2} \right) \mbox{sech}^2(\alpha), \;\;\; \alpha\geq 0.
\end{equation}
  Note that $\alpha\geq 0$ is a  constant parameter.   When $\alpha=0$, it is exactly the same as the traditional solitary waves of the KdV equation with a smooth crest.  

 In case of $\eta <  0$,  it holds $1-2 \eta/c > 1$ so that
\[   \mbox{tanh}^{-1}\left[ \sqrt{1-\frac{2 \eta}{c}} \right] \]
becomes a complex number  which has no physical meanings.   This is the reason why the solitary solutions of the KdV equation in case of $\eta<0$ was traditionally  neglected.  So, we must be very careful in this case.  

In case of $\eta\leq 0$,  write
\[  \sqrt{c - 2\eta} = \sqrt{c} \sqrt{1-\frac{2\eta}{c}} = \sqrt{c} \; z,    \]
where $z\geq 1$ and
\[ z^2  = 1-\frac{2 \eta}{c} \geq 1.    \] 
Then, we have 
\begin{equation}
 \eta = \frac{c}{2}(1-z^2), \;\;  d \eta = - c \; z \; d z.
\end{equation}
Since  $z \geq 1$, it holds
\begin{eqnarray}
&&\frac{d\eta}{\eta\sqrt{c-2\eta}} = \frac{1}{\sqrt{c}}\left( \frac{d z}{z-1}-\frac{d z}{z+1}\right)=d\left[\frac{1}{\sqrt{c}} \ln\left(\frac{z-1}{z+1}\right) \right].
\end{eqnarray}
Substituting it into (\ref{geq:KdV:4}) and integrating, we have
\begin{equation}
 \ln\left(\frac{z-1}{z+1}\right) = \pm \sqrt{c} \;  \xi - 2\beta,  \hspace{1.0cm}  \xi \geq 0,  \label{geq:KdV:5}
\end{equation}
where $\beta$ is a constant.   Since $z\geq 1$, it holds
\[    0 \leq  \frac{z-1}{z+1} < 1, \]
which gives
\[ \ln\left(\frac{z-1}{z+1}\right) < 0. \]
Thus, (\ref{geq:KdV:5})   holds for all $\xi\geq 0$  if  and only if
\begin{equation}
 \ln\left(\frac{z-1}{z+1}\right) = -(\sqrt{c} \; \xi +2 \beta), \;\;  \xi\geq 0, \;\; \beta>0,
\end{equation}
which gives
\[  z =  \mbox{coth}\left[ \frac{\sqrt{c}}{2} \; \xi  +\beta \right], \;\; \xi\geq 0, \;\; \beta>0.  \]
So, we have
\[
 \eta(\xi) = \frac{c}{2}(1-z^2)=-\left( \frac{c}{2} \right) \mbox{csch}^2\left[ \frac{\sqrt{c}}{2} \; \xi +\beta \right],
\;\; \xi\geq 0, \;\; \beta>0, \]
where csch denotes a hyperbolic cosecant function, and $\beta>0$ is a  constant. 
Using the symmetry condition (\ref{bc:symmetry}),  we have the solitary wave
\[
 \eta(\xi) = \frac{c}{2}(1-z^2)=-\left( \frac{c}{2} \right) \mbox{csch}^2\left[ \frac{\sqrt{c}}{2} \; |\xi| +\beta \right],
\;\; -\infty<\xi<+\infty, \;\; \beta>0, \]
which is valid in the {\em whole} interval.   

 Let  $d \leq \zeta(x,t) \leq 0$  denote the restriction  of $\zeta(x,t)$.  Then,  it holds 
\[   \eta(0) = -\left( \frac{c}{2} \right) \mbox{csch}^2  (\beta) \geq  d,  \]
which gives
\[    \beta \geq  \mbox{csch}^{-1}\sqrt{-\frac{2 d}{c}}. \]
Thus, we have the solitary waves  of the second kind:
\begin{equation}
\zeta(x,t) = -\frac{c}{2}  \mbox{csch}^2\left[ \frac{\sqrt{c}}{2}|x - c\;t - x_0| +\beta \right],  \label{zeta:new:negative}
\end{equation}
where
\[ \beta \geq  \mbox{csch}^{-1}\sqrt{-\frac{2 d}{c}}, \]
with the restriction  $d \leq \zeta(x,t) \leq 0$.    

The closed-form solutions (\ref{zeta:new:positive}) and (\ref{zeta:new:negative})  of the solitary waves of the first and second kind exactly satisfy the KdV equation (\ref{geq:KdV:original}) in the whole domain $-\infty < x<+\infty$ and $t\geq 0$,  but {\em except}  $x =  c \; t +x_0$ (corresponding to the wave crest).   This  is  rather  similar to the closed-form solution $u(x,t) = c \exp(-|x - c \; t|)$ of the CH equation (\ref{geq:CH}) when $\kappa=0$.  Thus,  the closed-form solutions (\ref{zeta:new:positive}) and (\ref{zeta:new:negative})  should be understood as weak solutions.   Besides,  the corresponding  Rankine-Hogoniot  jump  condition \cite{Rankine1870}  must be satisfied, so as to ensure that (\ref{zeta:new:positive}) and (\ref{zeta:new:negative})  have physical meanings.        

To give the corresponding Rankine-Hogoniot  jump  condition \cite{Rankine1870}  of the KdV equation (\ref{geq:KdV:original}), we rewrite it  in the flow
\begin{equation}
\zeta_t + [f(\zeta)]_x = 0,  \label{geq:jump}
\end{equation}
where $f(\zeta) = \zeta_{xx} + 3\zeta^2$.  Then,  $\zeta$  is a weak solution of (\ref{geq:jump}),  if
\begin{equation}
\int_{0}^{+\infty}\int_{-\infty}^{+\infty} \left[  \zeta \varphi_t + f(\zeta) \varphi_x  \right] dx dt + \int_{-\infty}^{+\infty} \zeta(x,0) \varphi(x,0) dx = 0 \label{def:weak}
\end{equation}
for all smooth functions $\varphi$ with compact support.   Besides,  there exists such a theorem that, if $\zeta$ is a weak solution of (\ref{geq:jump}) such that $\zeta$ is discontinuous across the curve
$x = \sigma(t)$ but $\zeta$ is smooth on either side of $x = \sigma(t)$, then $\zeta$ must satisfy the condition
 \begin{equation}
\sigma'(t) = \frac{f(\zeta^-) - f(\zeta^+)}{\zeta^- - \zeta^+},  \label{jump:general}
\end{equation}
across the curve of discontinuity, where $\zeta^-(x, t)$ is the limit of $\zeta$ approaching $(x,t)$ from the
left and $\zeta^+(x,t)$ is the limit of $\zeta$ approaching $(x, t)$ from the right.

Then,  due to the symmetry of the  progressive  peaked solitary waves  about the crest $x=c \; t + x_0$,  we have
\begin{eqnarray}
c &=& \frac{f(\zeta^-) - f(\zeta^+)}{\zeta^- - \zeta^+} = \frac{(\zeta^-)_{xx} -(\zeta^+)_{xx}+ 3(\zeta^-)^2-3(\zeta^+)^2}{\zeta^- - \zeta^+} \nonumber\\
&=& \frac{(\zeta^-)_{xx} -(\zeta^+)_{xx}}{\zeta^- - \zeta^+} + 3\left( \zeta^- + \zeta^+\right) \nonumber\\
&=& \frac{(\zeta^-)_{xxx} -(\zeta^+)_{xxx}}{(\zeta^-)_x - (\zeta^+)_x} + 3\left( \zeta^- + \zeta^+\right),   
\end{eqnarray}   
which provides us the so-called Rankine-Hogoniot  jump  condition of the KdV equation:
\begin{equation}
c = \frac{(\zeta^-)_{xxx} -(\zeta^+)_{xxx}}{(\zeta^-)_x - (\zeta^+)_x} + 3\left( \zeta^- + \zeta^+\right), \hspace{1.0cm} \mbox{as $x \to c \;t + x_0$.}
\label{jump:criterion}
\end{equation}

For the closed-form solution (\ref{zeta:new:positive}) at the crest, we have 
\begin{eqnarray}
\zeta^- &=&\zeta^+ = \frac{c}{2} \mbox{sech}^2(\alpha),\\
(\zeta^-)_x   & = &   \frac{c^{3/2}}{2}\mbox{sech}^2(\alpha) \mbox{tanh}(\alpha), \;\; (\zeta^+)_x = -(\zeta^-)_x,\\
(\zeta^-)_{xxx}   & = &   \frac{c^{5/2}}{4} \left[  \mbox{cosh}(2\alpha) -5\right] \mbox{sech}^4(\alpha) \mbox{tanh}(\alpha), \;\; 
 (\zeta^+)_{xxx} = -(\zeta^-)_{xxx}.
\end{eqnarray} 
Substituting them into (\ref{jump:criterion}), we have 
\begin{equation}
\frac{c}{2} \left[  \mbox{sech}^2(\alpha) +\mbox{sech}^2(\alpha)  \mbox{cosh}(2\alpha) -2  \right] = 0,
\end{equation}
which is an identity for not only $\alpha=0$ (corresponding to the traditional smooth solitary wave) but also {\em arbitrary} $\alpha > 0$ (to the peaked solitary wave)!  

Similarly, for the closed-form solution (\ref{zeta:new:negative}) at the crest, we have 
\begin{eqnarray}
\zeta^- &=&\zeta^+ = -\frac{c}{2} \mbox{csch}^2(\beta),\\
(\zeta^-)_x   & = &   -\frac{c^{3/2}}{2}\mbox{csch}^2(\beta) \mbox{coth}(\beta), \;\; (\zeta^+)_x = -(\zeta^-)_x,\\
(\zeta^-)_{xxx}   & = &   -\frac{c^{5/2}}{4} \left[  \mbox{cosh}(2\beta) +5\right] \mbox{csch}^4(\beta) \mbox{coth}(\beta), \;\; 
 (\zeta^+)_{xxx} = -(\zeta^-)_{xxx}.
\end{eqnarray} 
Substituting them into (\ref{jump:criterion}), we have 
\begin{equation}
\frac{c}{2} \left[  \mbox{csch}^2(\beta)   \;  \mbox{cosh}(2\beta)-\mbox{csch}^2(\beta)  - 2  \right] = 0,
\end{equation}
which is an identity for  {\em arbitrary}  constant $\beta > 0$.

 \begin{figure}[t]
\centering
\includegraphics[scale=0.5]{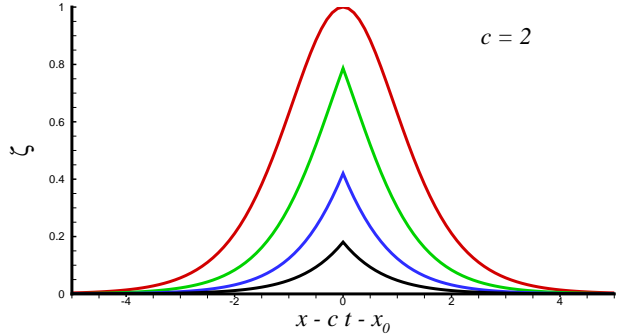}
\caption{ The solitary waves $\zeta(x,t)$ of the first kind of the KdV equation (\ref{geq:KdV:original}) with the same phase speed $c = 2$.   Red line: $\alpha=0$;  Green line: $\alpha=1/2$; Blue line: $\alpha =1$; Black line: $\alpha = 3/2$. }
\label{figure:KdV-1}
\end{figure}

 \begin{figure}[t]
\centering
\includegraphics[scale=0.5]{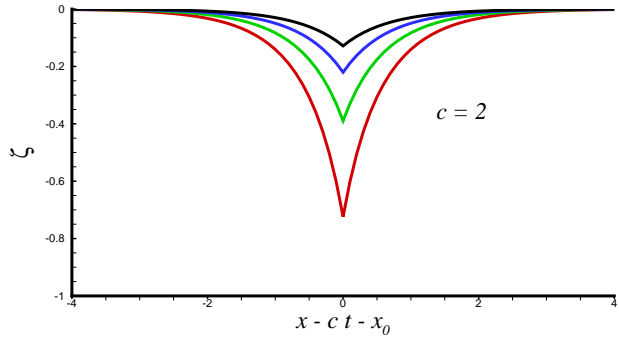}
\caption{ The solitary waves  of the second kind of the KdV equation (\ref{geq:KdV:original}) with the same phase speed  $c = 2$.   Red line: $\beta=1$;  Green line: $\beta=5/4$; Blue line: $\beta =3/2$; Black line: $\beta=7/4$. }
\label{figure:KdV-2}
\end{figure}

Therefore,  both of   (\ref{zeta:new:positive}) and (\ref{zeta:new:negative}) satisfy the corresponding Rankine-Hogoniot  jump  condition and should be   understood as weak  solutions of the KdV equation  (\ref{geq:KdV:original}).    It should be emphasized that the   traditional solitary wave (\ref{zeta:traditional}) with a smooth crest is  just only a special case of the solitary waves (\ref{zeta:new:positive})  of the first kind when $\alpha=0$.   Note that the solitary waves  (\ref{zeta:new:positive}) and (\ref{zeta:new:negative})  of the first and second kind of the KdV equation (\ref{geq:KdV:original}) have a peakon and an anti-peakon,  respectively.  Since $\alpha\geq 0$ and $\beta >0$ are arbitrary constant,  the phase speed of these peaked   solitary   waves  has nothing to do with the wave amplitude,  for example as shown in Figs.\ref{figure:KdV-1} and \ref{figure:KdV-2}.   This is quite different from the traditional  smooth solitary wave.    To the best of the author's knowledge,  such kind of peaked solitary waves  with  peakon or anti-peakon have  never  been  reported for the KdV equation.   All of these  reveal the novelty of the peaked solitary waves (\ref{zeta:new:positive}) and (\ref{zeta:new:negative}).  

\section{Peaked solitary waves of modified KdV equation}%

Similarly,  the modified KdV equation  \cite{mKdV, Zabusky1967}
\begin{equation}
\zeta_t + \zeta_{xxx} \pm 6  \zeta^2  \zeta_x =0   \label{geq:mKdV}
\end{equation}
has the two kinds of    solitary waves
\begin{equation}
\zeta(x,t) = \pm \frac{2 c }{e^{\sqrt{c} |x- c\; t-x_0|+\alpha}\pm c \; e^{-\sqrt{c} |x- c\; t-x_0|-\alpha} }, \label{u:mKdV}
\end{equation}
where $c$ is the phase speed and $\alpha\geq 0$ is a  constant.   

Note that (\ref{jump:general}) holds for  (\ref{geq:jump}) in general.   Now, we have $f(\zeta) = \zeta_{xx} \pm 2  \zeta^3$.   And the corresponding Rankine-Hogoniot  jump  condition reads
\begin{eqnarray}
c &=& \frac{f(\zeta^-)-f(\zeta^+)}{\zeta^- - \zeta^+} \nonumber\\
&=&\frac{(\zeta^-)_{xx} -(\zeta^+)_{xx} \pm  2\left[ (\zeta^-)^3 - (\zeta^+)^3\right] }{\zeta^- - \zeta^+} \nonumber\\
&=& \frac{(\zeta^-)_{xx} -(\zeta^+)_{xx}}{\zeta^- - \zeta^+}  \pm  2\left[  (\zeta^-)^2 +\zeta^-\zeta^+ + (\zeta^+)^2\right] \nonumber\\
&=& \frac{(\zeta^-)_{xxx} -(\zeta^+)_{xxx}}{(\zeta^-)_x - (\zeta^+)_x} \pm  2\left[  (\zeta^-)^2 +\zeta^-\zeta^+ + (\zeta^+)^2\right],    
\end{eqnarray}
which provides us the Rankine-Hogoniot  jump  condition of the modified KdV equation
\begin{equation}
c = \frac{(\zeta^-)_{xxx} -(\zeta^+)_{xxx}}{(\zeta^-)_x - (\zeta^+)_x} \pm  2\left[  (\zeta^-)^2 +\zeta^-\zeta^+ + (\zeta^+)^2\right], \hspace{1.0cm} \mbox{as $x \to c \; t + x_0$}.  \label{jump:mKdV}
\end{equation}
Using  the closed-form solution  (\ref{u:mKdV}), it is found that the above condition is satisfied for {\em arbitrary} constant $\alpha\geq 0$.    Therefore,  the closed-form solution (\ref{u:mKdV})   exactly  satisfies  the corresponding  Rankine-Hogoniot  jump  condition of the modified KdV equation (\ref{geq:mKdV})  and should  be  understood as a weak solution.

\section{Peaked solitary waves of  the BBM equation}%

Similarly,  the  BBM equation \cite{Benjamin1972}
\begin{equation}
   \zeta_t + \zeta_x + \zeta \; \zeta_x - \zeta_{xxt} = 0  \label{geq:BBM}
   \end{equation}
has the solitary waves of the first kind
\begin{equation}
\zeta(x,t) = 3(c-1) \mbox{sech}^2\left[\frac{\sqrt{1-c^{-1}}}{2}\left|\xi\right|+\alpha \right], \alpha\geq 0, \label{BBM:1st} 
\end{equation}
and the solitary waves of the second kind
\begin{equation}
\zeta(x,t) = -3(c-1) \mbox{csch}^2\left[\frac{\sqrt{1-c^{-1}}}{2}\left|\xi\right|+\beta \right], \beta > 0, \label{BBM:2nd}
\end{equation}
where $\xi = x - c \; t - x_0$,  $c$ is the phase speed,  and $\alpha\geq 0$, $\beta>0$ are   constants. 

Now, we have $f(\zeta) = \zeta + \zeta^2/2 -\zeta_{xt}$.   And the corresponding Rankine-Hogoniot  jump  condition reads
\begin{eqnarray}
c &=& \frac{f(\zeta^-)-f(\zeta^+)}{\zeta^- - \zeta^+} \nonumber\\
&=&\frac{(\zeta^-)-(\zeta^+) + \left[(\zeta^-)^2-(\zeta^+)^2\right]/2- \left[ (\zeta^-)_{xt} -(\zeta^+)_{xt}\right] }{\zeta^- - \zeta^+} \nonumber\\    
&=& 1 + \frac{1}{2}\left(\zeta^- + \zeta^+\right)-\frac{(\zeta^-)_{xxt} -(\zeta^+)_{xxt}}{(\zeta^-)_x - (\zeta^+)_x},
\end{eqnarray}
which provides us the Rankine-Hogoniot  jump  condition of the MMB equation
\begin{equation}
c = 1 + \frac{1}{2}\left(\zeta^-+\zeta^+\right)-\frac{(\zeta^-)_{xxt} -(\zeta^+)_{xxt}}{(\zeta^-)_x - (\zeta^+)_x}, \;\;\; \mbox{as  $x \to  c\; t + x_0$}.\label{jump:BBM}
\end{equation}

For the  solitary waves (\ref{BBM:1st}) of the first kind at the crest, we have 
\begin{eqnarray}
\zeta^- &=&  3(c-1) \mbox{sech}^2(\alpha), \\
(\zeta^-)_x &=& 3 c (1-c^{-1})^{3/2}  \mbox{sech}^2(\alpha)\; \mbox{tanh}(\alpha),\;\; \\
 (\zeta^-)_{xxt} &=& -\frac{3}{4} c^2 (1-c^{-1})^{5/2} \mbox{sech}^5(\alpha) \left[ \mbox{sinh}(3\alpha) -11 \mbox{sinh}(\alpha)\right], 
\end{eqnarray} 
and 
\[ \zeta^+ = \zeta^-, \;\;  (\zeta^+)_x =-(\zeta^-)_x,  \;\; (\zeta^+)_{xxt} =-(\zeta^-)_{xxt}.  \]
Substituting all of these into (\ref{jump:BBM}),  it is found that  (\ref{BBM:1st})  satisfies  the  Rankine-Hogoniot  jump  condition  (\ref{jump:BBM}) for {\em arbitrary}  constant $\alpha\geq 0$.    Similarly,  (\ref{BBM:2nd})  satisfies  the  Rankine-Hogoniot  jump  condition  (\ref{jump:BBM}) for {\em arbitrary}  constant $\beta>0$.  Therefore, both of   (\ref{BBM:1st})   and (\ref{BBM:2nd}) should be understood  as  weak solutions of the BBM equation (\ref{geq:BBM}).   

\section{Peaked solitary waves of Boussinesq equation}%

  In addition, the Boussinesq equation  \cite{Boussinesq1872}
\begin{equation}
\frac{\partial^2 \zeta}{\partial t^2} - g h
\frac{\partial^2 \zeta}{\partial x^2} -g h
\frac{\partial^2 }{\partial x^2}\left(\frac{3\zeta^2}{2h}  + \frac{h^2}{3}\frac{\partial^2 \zeta}{\partial x^2}
 \right) = 0  \label{geq:Boussinesq}
\end{equation}
has the solitary waves of the first kind
\begin{equation}
\zeta = h\left(\frac{c^2}{g h}-1 \right)\mbox{sech}^2\left[\frac{\sqrt{3}}{2 h}\sqrt{\frac{c^2}{g h}-1} \; \left|\xi\right| +\alpha \right], \label{Boussinesq:1st}
\end{equation} 
and the solitary waves of the second kind
\begin{equation}
\zeta = -h\left(\frac{c^2}{g h}-1 \right)\mbox{csch}^2\left[\frac{\sqrt{3}}{2 h}\sqrt{\frac{c^2}{g h}-1} \;  \left|\xi\right| +\beta \right], \label{Boussinesq:2nd}
\end{equation}
where $\xi = x-c \; t - x_0$,  $h$ denotes the water depth, $g$ the acceleration of gravity, $c$ the phase speed of wave,  $\alpha\geq 0$, $\beta>0$ are  constant,  respectively.   

For progressive solitary wave, we have 
\[   \frac{\partial }{\partial x} = -\frac{1}{c} \frac{\partial }{\partial t}.    \] 
Then,   (\ref{geq:Boussinesq}) can be rewritten in the form
\begin{equation}
\frac{\partial^2 \zeta}{\partial t^2} +\frac{g h}{c}
\frac{\partial^2 \zeta}{\partial x \partial t} +\frac{g h}{c}
\frac{\partial^2 }{\partial x \partial t}\left(\frac{3\zeta^2}{2h}  + \frac{h^2}{3}\frac{\partial^2 \zeta}{\partial x^2}
 \right) = 0.  \label{geq:Boussinesq:2}
\end{equation}
Integrating it with respect to $t$ and using the boundary condition $\zeta(\pm \infty)=0$, we have 
\begin{equation}
\zeta_t + \frac{g h}{c} \left[\zeta + \frac{3\zeta^2}{2h}  + \frac{h^2}{3} \zeta_{xx} \right]_x = 0.
\end{equation}
Thus, we have here
\[   f(\zeta) =  \frac{g h}{c} \left(\zeta + \frac{3\zeta^2}{2h}  + \frac{h^2}{3} \zeta_{xx} \right).  \]
According to (\ref{jump:general}),  we have
\begin{eqnarray}
c  &=&  \frac{f(\zeta^-) - f(\zeta^+)}{\zeta^- - \zeta^+}   \nonumber\\
&=& \left( \frac{g h}{c} \right)\frac{(\zeta^- - \zeta^+) + \frac{3}{2h}\left[(\zeta^-)^2 - (\zeta^+)^2\right]  +\frac{h^2}{3}\left[ (\zeta^-)_{xx} - (\zeta^+)_{xx}\right] }{\zeta^- - \zeta^+} \nonumber\\
&=& \left( \frac{g h}{c} \right) \left[ 1 +  \frac{3}{2h} \left( \zeta^- + \zeta^+\right) +\left(\frac{h^2}{3}\right)
\frac{ (\zeta^-)_{xx} - (\zeta^+)_{xx}}{\zeta^- - \zeta^+} 
     \right] \nonumber\\
 &=& \left( \frac{g h}{c} \right) \left[ 1 +  \frac{3}{2h} \left( \zeta^- + \zeta^+\right) +\left(\frac{h^2}{3}\right)
\frac{ (\zeta^-)_{xxx} - (\zeta^+)_{xxx}}{(\zeta^-)_x - (\zeta^+)_x} 
     \right],   
\end{eqnarray}
which provides us the Rankine-Hogoniot  jump  condition of the Boussinesq  equation
\begin{equation}
\frac{c^2}{g h} =   1 +  \frac{3}{2h} \left( \zeta^- + \zeta^+\right) +\left(\frac{h^2}{3}\right)
\frac{ (\zeta^-)_{xxx} - (\zeta^+)_{xxx}}{(\zeta^-)_x - (\zeta^+)_x}, \hspace{1.0cm} \mbox{as $x\to c\; t + x_0 $}.\label{jump:Boussinesq}
\end{equation}

For the peaked solitary wave (\ref{Boussinesq:1st}), we have at the crest that 
\begin{eqnarray}
\zeta^- &=& h \left( \frac{c^2}{g h}-1\right)\mbox{sech}^2 (\alpha),\\
(\zeta^-)_x &=& \sqrt{3} \left( \frac{c^2}{g h}-1\right)^{3/2} \mbox{sech}^2(\alpha) \; \mbox{tanh}(\alpha), \\
(\zeta^-)_{xxx} &=& \frac{3\sqrt{3}}{4 h^2}  \left( \frac{c^2}{g h}-1\right)^{5/2} \mbox{sech}^2(\alpha)
\left[\mbox{sinh}(3\alpha) -11\mbox{sinh}(\alpha) \right],
\end{eqnarray}
and  
\[   (\zeta^+)_{x} =-(\zeta^+)_{x},\;\;  (\zeta^+)_{xxx} =-(\zeta^+)_{xxx}.   \]
Substituting all of these into (\ref{jump:Boussinesq}), we have 
\begin{equation}
\frac{1}{4}\left( \frac{c^2}{g h}-1\right) \left[  \mbox{sech}^2(\alpha)+\mbox{sech}^2(\alpha) \; \mbox{csch}(\alpha) \; \mbox{sinh}(3\alpha)-4  \right] = 0,
\end{equation}
which is an identity for  {\em arbitrary}  constant $\alpha \geq 0$.    Similarly, it is found that (\ref{Boussinesq:2nd}) satisfies the  Rankine-Hogoniot  jump  condition (\ref{jump:Boussinesq}) for {\em arbitrary} constant $\beta>0$.   Thus,   the peaked solitary waves  (\ref{Boussinesq:1st}) and (\ref{Boussinesq:2nd}) should be understood as weak  solutions  of  Boussinesq equation (\ref{geq:Boussinesq}).

\section{Conclusion and discussion}

In this article, we give, for the first time,  the closed-form solutions  of the peaked solitary waves of  the KdV equation \cite{KdV},  the modified KdV equation \cite{mKdV, Zabusky1967}, the BBM equation \cite{Benjamin1972}, and Boussinesq equation \cite{Boussinesq1872}, respectively.   All of them  exactly satisfy the corresponding  Rankine-Hogoniot   jump  condition for {\em arbitrary} constant $\alpha\geq 0$ or $\beta >0$.   Note that the 1st-derivative  $\zeta_x$ of the elevation has a jump at the crest.  Obviously, for all peaked solitary waves found in this article,  $\zeta_x >0$  on the left of the crest, but $\zeta_x<0$ on the right of the crest, respectively.  In other words, $\zeta_x^- > \zeta_x^+$.  So,  these  peaked  solutions also satisfy the so-called entropy condition.   Thus,  they could be understood as weak solutions.     

Note that the traditional  smooth solitary waves are just special cases of the peaked solitary waves when $\alpha=0$.   So, these weak solutions are more general.  Besides,  unlike the smooth solitary waves whose phase speed strongly depends   upon wave amplitude, the peaked solitary waves (when $\alpha>0$ or $\beta>0$) have {\em nothing} to do with the wave amplitude, for example as shown in  Figs.\ref{figure:KdV-1} and \ref{figure:KdV-2}.   This is quite different from the traditional solitary waves.    In addition, the  solitary waves  with  negative elevation such as shown in Fig.~\ref{figure:KdV-2}  have never been reported for these mainstream models of shallow waves.  All of these show the  novelty of these peaked solitary waves.   If  these  peaked  waves  as  weak solutions indeed exist and have physical meanings, then nearly all  mainstream   models for  shallow  water  waves,  including the KdV equation \cite{KdV}, the modified KdV equation \cite{mKdV, Zabusky1967}, the BBM equation \cite{Benjamin1972}, and Boussinesq equation \cite{Boussinesq1872} and  the CH equation \cite{Camassa1993PRL},   have  the  peaked  solitary  waves, no matter whether or not they are integral and admit breaking-wave solutions.  Thus,   the peaked solitary waves might  be a common property of shallow water wave models.  As shown currently by Liao \cite{Liao-arXiv-PeakedWave},  even the exactly nonlinear water wave equations also admit the peaked solitary waves:  this might reveal the origin of the peaked solitary waves in shallow water predicted by these mainstream  models.      
          
Certainly,  further investigations on these peaked solitary waves are needed, especially the stability of them, the interactions between multiple peaked solitary waves, and so on.  Note that these peaked solitary waves have discontinuous 1st derivative at crest so that their higher derivatives at crest tend to infinity  and  the perturbation theory does not work.   Thus,  strictly speaking,   the validity of the KdV equation \cite{KdV},  the modified KdV equation \cite{mKdV, Zabusky1967}, the BBM equation  \cite{Benjamin1972}, and Boussinesq equation \cite{Boussinesq1872}  should be verified  carefully  in the meaning of the weak solution (\ref{def:weak}).    
%Note that  the Rankine-Hogoniot   jump  condition and  entropy condition are only  necessary conditions for a weak solution.   So, there are some open questions.  For example, how do these peaked solitary waves come into being?  How  to  directly  prove  that  they  are  indeed weak solutions  satisfying  (\ref{def:weak})?  

Note that  the peaked solitary waves  have never been observed experimentally and in practice.   Obviously, it is more difficult to create and remain such kinds of peaked solitary waves than the traditional smooth ones.  So, it is an challenging work to observe these peaked solitary waves  in experiments.      

It should be emphasized that  the discontinuity and/or  singularity exist widely in natural phenomena, such as dam break in hydrodynamics, shock waves in aerodynamics,  black holes described by general relativity  and so on.  Indeed, the discontinuity and/or singularity are rather difficult to handle by traditional methods.  But, the discontinuity and/or singularity  can greatly enrich and deepen  our understandings about the real world, and therefore should not be  evaded easily.     

\section*{Acknowledgement}

Thanks to Prof. Y.G. Wang (Dept. of Mathematics, SHJT) for introducing me the concept of the Rankine-Hogoniot  jump  condition of weak solution.  Thanks to Prof. C.C. Mei  (MIT) for mentioning the validity of perturbation method to the mainstream  models of shallow water waves in case of peaked crest.   This work is partly supported by the State Key Lab of Ocean Engineering (Approval No. GKZD010056-6) and the National Natural Science Foundation of China.

\bibliographystyle{unsrt}

\end{document}